\documentclass{article}
\usepackage{spconf,amsmath,graphicx}
\usepackage{amssymb}
\usepackage{cite}
\usepackage{upgreek}
\usepackage{tipa}

\title{Geometry-aware DoA Estimation using a Deep Neural Network with mixed-data input features}
\name{
    Ulrik Kowalk$^{1}$\thanks{This work was funded by the German Federal Ministry of Education and
Research under the funding program "Forschung an Fachhochschulen", Project ID: 13FH666IB6.}, 
    Simon Doclo$^{2}$, 
    and Joerg Bitzer$^{1}$
    }
\address{
    $^{1}$Jade University of Applied Sciences, Institute for Hearing Technology and Audiology, Oldenburg, Germany\\ 
    $^{2}$University of Oldenburg, Department of Medical Physics and Acoustics and\\Cluster of Excellence Hearing4all, Oldenburg, Germany
    }
\begin{document}
\ninept
\maketitle
\begin{abstract}
Unlike model-based direction of arrival (DoA) estimation algorithms, supervised learning-based DoA estimation algorithms based on deep neural networks (DNNs) are usually trained for one specific microphone array geometry, resulting in poor performance when applied to a different array geometry. 
In this paper we illustrate the fundamental difference between supervised learning-based and model-based algorithms leading to this sensitivity.
Aiming at designing a supervised learning-based DoA estimation algorithm that generalizes well to different array geometries, in this paper we propose a geometry-aware DoA estimation algorithm. The algorithm uses a fully connected DNN and takes mixed data as input features, namely the time lags maximizing the generalized cross-correlation with phase transform and the microphone coordinates, which are assumed to be known.
Experimental results for a reverberant scenario demonstrate the flexibility of the proposed algorithm towards different array geometries and show that the proposed algorithm outperforms model-based algorithms such as steered response power with phase transform.
\end{abstract}
\section{Introduction}
\label{sec:introduction}
%
\vspace*{-2mm}
%
%
Estimating the direction of arrival (DoA) of an acoustic source is required in many speech communication applications, e.g. to steer a beamformer in hearing aids or to steer a camera in a teleconferencing system.\cite{huang2000microphone,doclo2010acoustic}
Most DoA estimation techniques use multiple microphone signals that are recorded by a microphone array with a known geometry.\cite{huang2008time,madhu2008acoustic}
Apart from classical model-based approaches that exploit time differences at the microphones directly, e.g., the generalized cross-correlation with phase transform (GCC-PHAT)\cite{knapp1976generalized,1457990}, steered response power with phase transform (SRP-PHAT)\cite{dibiase2000high}, and subspace-based methods such as multiple signal classification (MUSIC)\cite{schmidt1986multiple}, in the last years many supervised learning-based algorithms based on deep neural networks (DNNs) have been proposed for DoA estimation\cite{grumiaux2022survey,wang2018robust,8651493,zhang2019robust,mack2022signal,krause2021data}. 

While model-based DoA estimation algorithms exploit different signal properties, such as time difference of arrival\cite{knapp1976generalized} or the covariance matrix estimated from the microphone signals\cite{schmidt1986multiple}, at their core is the computation of an analytic function that incorporates the microphone array geometry. 
For example in SRP-PHAT\cite{dibiase2000high,johansson2002speaker} the acoustic power is sampled at candidate DoAs by an acoustic beamformer whose steering vector depends on the microphone array geometry. Similarly, the MUSIC pseudo-spectrum\cite{schmidt1986multiple} is based on the orthogonal projection of the noise subspace of the covariance matrix of the microphone signals with an array geometry-dependent steering vector.

\begin{figure}[t]
    \centering
    \includegraphics[width=\linewidth]{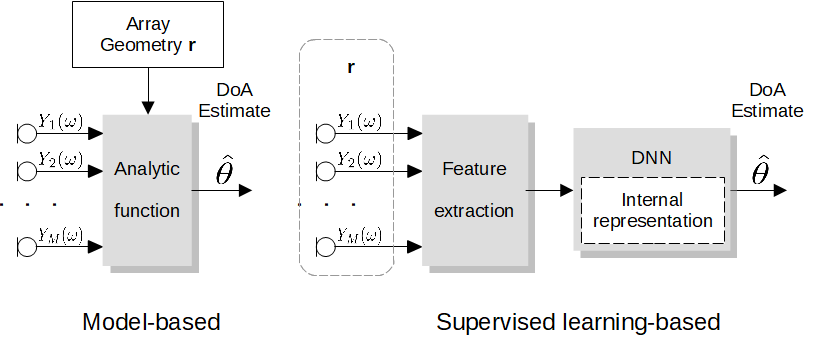}
    \caption{Model-based DoA estimation approaches: Computation of an analytic function based on the array geometry \textbf{r}. Supervised learning-based approaches: DoA estimation using an internal geometry representation, which depends on the array geometry assumed during training} 
    \label{fig:Comparison_model_dnn}
    \vspace*{-3mm}
\end{figure}

The preceding observations do not apply to supervised learning-based DoA estimation algorithms, where a DNN learns the relationship between input features and the DoA. These data-driven systems are trained using either microphone array recordings (simulated or real-world) or features that are extracted from such recordings. As a consequence, all training data is implicitly based on the underlying array geometry\cite{zhang21e_interspeech}, of which the DNN compiles some form of internal representation.
Without explicit information on the array geometry, the DNN is only able to learn the relationship between input features and DoA provided that all data originates from the same geometry. This holds for fully connected deep neural networks (FC-DNNs) as well as for convolutional neural networks (CNNs), in which the first layers perform a convolution operation on the input features.
If this geometry assumption is not satisfied at inference, e.g., a DNN trained for one array geometry is applied to another array geometry, the DoA estimation performance may be substantially degraded unless a retraining step is performed, requiring a completely new data set\cite{9746876} besides processing time and power. 
Fig.\,\ref{fig:Comparison_model_dnn} illustrates the general dependency of model-based and supervised learning-based DoA estimation algorithms on the array geometry, i.e. model-based algorithms require knowledge of the microphone array geometry to compute an analytic function, whereas supervised learning-based techniques operate using an internal geometry representation.
In this paper we present a feasibility study on geometry-aware supervised learning-based DoA estimation. We propose a deep neural network that takes as input two separate types of independent data, namely the time lags maximizing the GCC-PHAT and the microphone array geometry, which is assumed to be perfectly known.
Based on two experiments we demonstrate that the proposed geometry-aware DNN is flexible towards different array geometries and outperforms state-of-the-art model-based algorithms.
%

\section{Model-based DoA estimation}
\label{sec:signal_model}

We assume an array of $M$ omnidirectional microphones with a known geometry \textbf{r} capturing a single static acoustic source in a reverberant environment where some background noise is present. In the frequency domain, the signal at the $m$-th microphone is composed of 
\begin{equation}
    Y_{m}(\omega)=S(\omega)H_{m}(\omega)+V_{m}(\omega),
\end{equation}
where $S(\omega)$ denotes the source signal, $H_{m}(\omega)$ denotes the acoustic transfer function between the acoustic source and the $m$-th microphone, and $V_{m}(\omega)$ denotes the additive noise component at the $m$-th microphone. 

Based on this signal model, model-based algorithms estimate the DoA of the acoustic source by computing an analytic function incorporating knowledge of the microphone array geometry.
For example, SRP-PHAT computes an acoustic power map for candidate DoAs $\theta$ using an acoustic beamformer whose steering vector is based directly on the microphone array geometry \textbf{r} as
\vspace*{-1mm}
\begin{equation}
    P(\theta, \textbf{r}) = 2\pi \sum\limits_{k=1}^{M}\sum\limits_{l=1}^{M}\int\limits_{-\infty}^{\infty} \Gamma_{k,l}(\omega) e^{j\omega\tau_{k,l}(\theta)}d\omega,
\end{equation}
%
where $\boldsymbol{\Gamma}_{k,l}$ denotes the frequency-domain representation of the GCC-PHAT,  $\boldsymbol{\upgamma}_{k,l}$, between microphones $k,l$, which is defined as
\begin{equation}
    \boldsymbol{\upgamma}_{k,l} = \mathcal{F}^{-1} \left\{ \frac{Y_{k}(\omega) \cdot Y^{*}_{l} (\omega)}{|Y_{k}(\omega) \cdot Y^{*}_{l} (\omega)|} \right\},
    \label{eq:GCC-PHAT_03}
\end{equation}
where $\mathcal{F}^{-1}$ denotes the inverse Fourier transform and $(\cdot)^{*}$ denotes complex conjugation. The DoA-specific time delay $\tau_{k,l}(\theta)$ between microphones $k,l$ is determined by their coordinates, i.e.
\begin{equation}
    \tau_{k,l}(\theta) = \frac{| \textbf{r}_{k} - \textbf{r}_{l} | \sin{\theta}}{c},
\end{equation}
where $c$ is the speed of sound, indicating the direct dependence of $P$ on $\textbf{r}$. The MUSIC algorithm\cite{schmidt1986multiple} computes the noise subspace from the covariance matrix of the microphone signals and establishes the so-called MUSIC pseudo-spectrum $P(\theta,\textbf{r})$  as
\begin{equation}
   P(\theta,\textbf{r}) = \frac{1}{|| \textbf{a}^{H}(\theta,\textbf{r})\textbf{E}_{N}\textbf{E}_{N}^{H}\textbf{a}(\theta,\textbf{r})||}, 
\end{equation}
where $\textbf{E}_{N }$ denotes the noise subspace and ${\bf a}(\theta,\textbf{r})$ is an array geometry-dependent steering vector, again indicating a direct dependence of $P$ on \textbf{r} and showing the flexibility of model-based algorithms towards different microphone array geometries.

\section{Supervised learning for DoA estimation}
\label{sec:geometry-aware_sl}

In this section we discuss three supervised learning-based DoA estimation algorithms, which share the same FC-DNN architecture (Section\,\ref{sec:dnn_architecture}). After describing the conventional GCC-PHAT input features in Section\,\ref{sec:fcfull}, in Section\,\ref{sec:fcmax} we introduce a reduced set of GCC-PHAT-based input features. In Section\,\ref{sec:geometry-aware}, we propose a geometry-aware DoA estimation algorithm, which uses the microphone geometry as additional input features.

\subsection{DNN Architecture}
\label{sec:dnn_architecture}

In this study, the DoA estimation problem is formulated as a multi-class classification task, with $C$ DoA classes $\boldsymbol{\theta}$ that span the 360° azimuth range. For each signal frame the class maximizing the posterior probability map \textit{\textbf{P}} in the output layer is determined by     
\begin{equation}
    I = \underset{i\ \in\ C}{\arg\max}\ P_{i}, 
\end{equation}
with the DoA estimate given by $\hat{\theta}=\theta_{I}$.
Because in most cases the posterior probability is not concentrated in a single class, a better estimate is obtained through parabolic interpolation\cite{smith1987parshl} incorporating the three DoA classes centered around the maximum. For signals consisting of multiple frames, the global DoA estimate $\tilde{\theta}$ is calculated as the median value over all frames.

\begin{figure}[t]
    \centering
    \includegraphics[width=\linewidth]{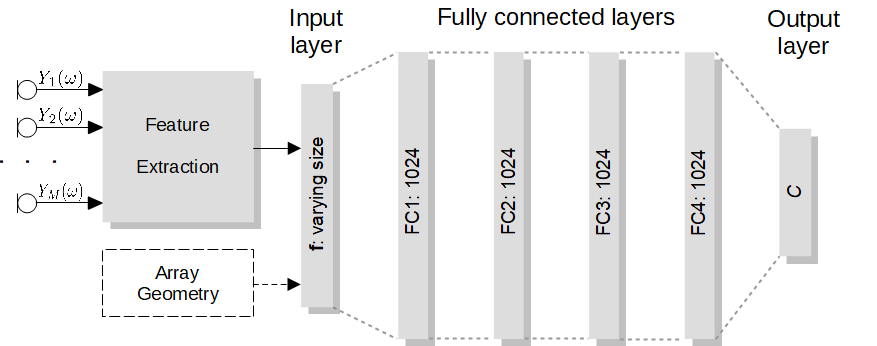}
    \caption{Universal FC-DNN architecture comprising an input layer of varying size, depending on the algorithm-specific input features, four fully connected (FC) layers, and an output layer with $C$ neurons.} 
    \label{fig:geometry-aware}
\end{figure}

For all supervised learning-based algorithms we consider an FC-DNN architecture (see Fig.\,\ref{fig:geometry-aware}) consisting of an input layer, four fully connected hidden layers with 1024 neurons each, followed by an output layer with \textit{C} neurons. Every fully connected layer comprises a 20\,\% dropout stage and is activated by a ReLU function. Because each of the FC-DNNs uses different input features, the sizes of their input layers are different, as indicated in Fig.\,\ref{fig:geometry-aware}.

\subsection{FC-DNN with GCC-PHAT (FC$_{full}$)}
\label{sec:fcfull}

The first FC-DNN uses a discrete time version of the GCC-PHAT in (\ref{eq:GCC-PHAT_03}) as input feature. GCC-PHAT is a representation of the time differences between microphone signals and has been shown to be robust against reverberation and noise due to its phase transform property\cite{zhang2008does}. 
For practical reasons we constrain the discrete time lag $\tau^{\delta}$ to the interval [$-\tau^{\delta}_{max}$,$\tau^{\delta}_{max}-$1], where $\tau^{\delta}_{max}$ corresponds to the largest possible inter-microphone time delay in samples, i.e.
\vspace*{-1mm}
\begin{equation}
    \tau^{\delta}_{max} = \bigg\lceil \frac{r_{max} \cdot f_{s}}{c}\bigg\rceil + \eta,
\end{equation}
where $r_{max}$ denotes the largest inter-microphone distance within the array, $f_{s}$ denotes the sampling rate, \textit{c} denotes the speed of sound, $\lceil\cdot\rceil$ indicates the operation of rounding up, and $\eta$ denotes an additional margin. By concatenating the discrete GCC-PHAT vectors for all non-redundant microphone pairs, the final feature vector is constructed as
\begin{equation}
    \boxed{
        \textbf{f}_{full} = [\boldsymbol{\upgamma}^{\delta}_{1,2},\boldsymbol{\upgamma}^{\delta}_{1,3},\dots,\boldsymbol{\upgamma}^{\delta}_{M-1,M}]
    }
    \label{eq:feature_vector_full}
\end{equation}
Because this FC-DNN uses the full GCC-PHAT feature vector in (\ref{eq:feature_vector_full}) as input feature, it is referred to as FC$_{full}$. The size of the input layer is equal to $M(M-1)/2 \cdot 2\tau^{\delta}_{max}$, i.e., the number of non-redundant microphone pairs multiplied by the width of each GCC-PHAT.

\subsection{FC-DNN with GCC-PHAT maximum locations (FC$_{max}$)}
\label{sec:fcmax}

The second FC-DNN uses a reduced feature set of GCC-PHAT-based features as input feature, where we propose to only use the location of the maximum in each GCC-PHAT vector. This is performed for three reasons. First, the size of the feature set is drastically reduced. Second, this excludes local maxima in the GCC-PHAT, e.g. arising from room reflections. Third, it can be argued that the location of the maximum is the most important piece of information in GCC-PHAT, where we would like the DNN to focus on. We first determine the discrete time lag that maximizes the GCC-PHAT using
\vspace*{-2mm}
\begin{equation}
    d_{k,l} = \underset{\tau^{\delta}}{\arg\max}\ \boldsymbol{\upgamma}^{\delta}_{k,l}.
    \vspace*{-1mm}
\end{equation}
Since the true maximum is most likely situated between two discrete time lags, we then determine the interpolated time lag $\tilde{d}_{k,l}$ by applying parabolic interpolation between $d_{k,l}$-1 and $d_{k,l}$+1. The final feature vector is constructed by concatenating all estimated time lags as
\begin{equation}
    \boxed{
        \textbf{f}_{max} = [\tilde{d}_{1,2},\tilde{d}_{1,3},\ldots,\tilde{d}_{M-1,M}]
    }
    \label{eq:feature_vector_max}
    \vspace*{1mm}
\end{equation}
Because this DNN uses the locations of GCC-PHAT maxima, it is referred to as FC$_{max}$.
The size of the input layer is equal to\linebreak$M(M-1)/2$, i.e., the number of non-redundant microphone pairs.

\vspace*{-1mm}
\subsection{Geometry-aware FC-DNN (FC$_{GA}$)}
\label{sec:geometry-aware}

Whereas the FC-DNNs in the previous sections only use signal-based input features, in this section we propose a geometry-aware FC-DNN, which uses a combination of two types of independent data as input feature: the GCC-PHAT maximum locations, $\textbf{f}_{max}$ in (\ref{eq:feature_vector_max}), and the microphone coordinates $\textbf{f}_{r}$, separated into their $x$ and $y$ components, i.e.
\vspace*{-1mm}
\begin{equation}
    \textbf{f}_{r} = [x_{1},\dots,x_{M},y_{1},\dots,y_{M}]
\end{equation} 
Both parts, maximum locations and array coordinates, are then concatenated to form the final feature set:
%
\begin{equation}
    \boxed{
        \textbf{f} = [\textbf{f}_{max}, \textbf{f}_{r}]
    }
\end{equation}
It should be noted that experiments have shown that the combination of GCC-PHAT maximum locations and microphone coordinates produces more accurate and robust estimates than the combination of the full GCC-PHAT vector and microphone coordinates. 
Since this DNN is geometry-aware, it is referred to as FC$_{GA}$. The size of the input layer is equal to $M(M-1)/2+2M$, i.e., the number of non-redundant microphone pairs plus the $x$- and $y$-coordinates of the microphones.

\vspace*{-2mm}
\section{Experimental Evaluation}
\label{sec:experimental_validation}

In this section the performance of the proposed geometry-aware DNN is validated and compared to several baseline algorithms for different acoustic conditions. 
In Section\,\ref{sec:acoustic_setup} the considered acoustic setup is presented, while Section\,\ref{sec:training} describes the training procedure. The performance metrics are introduced in Section\,\ref{sec:performance_metrics}, and the evaluation results are presented in Section\,\ref{sec:evaluation}.  
In addition to FC$_{full}$ and FC$_{max}$ as baseline algorithm we consider the CNN-based algorithm from \cite{8651493}, which uses raw signal phases as input feature. We further consider SRP-PHAT and MUSIC, both implemented in \cite{scheibler2018pyroomacoustics}, as baseline model-based algorithms as well.

\vspace*{-1mm}
\subsection{Acoustic Setup and Algorithm Parameters}
\label{sec:acoustic_setup}

\begin{table}[t]
    \centering
    \begin{tabular}{|l | l|}
        \hline
         Room dimensions: & [9.0, 5.0, 3.0]\,m $\pm$ [1.0, 1.0, 0.5]\,m \\ 
         Array position: & [4.5, 2.5, 1.5]\,m $\pm$ [0.5, 0.5, 0.5]\,m \\  
         Source distance: & 1.0 - 3.0\,m [within boundaries]\\
         Source direction: & 0°\ :\ 5°\ :\ 355°\\
         $T_{60}$: & 0.13\,s - 1.0\,s  \\
         SNR: & 0 - 30\,dB\\
         \hline
        \end{tabular}
        \vspace*{-2mm}
    \caption{\label{tab:Variations}Acoustic simulation parameters}
    \vspace{-4mm}
\end{table}
%


For the evaluation we consider a 2-dimensional arc-shaped microphone array comprising $M$=5 microphones with a width of 0.4\,m and a depth of about 0.15\,m, as shown in Fig.\,\ref{fig:array_shape}. 
The microphone array is situated inside a rectangular room and records the signal from a single acoustic source and noise. 
For both training and evaluation we utilize data consisting of 50\% speech and 50\% white noise as suggested in \cite{krause2021data}. Speech data is taken from the ``clean'' section of the \textit{LibriSpeech} corpus\cite{panayotov2015librispeech}. Directional cues are simulated by convolving monophonic signals with room impulse responses (RIRs) generated using the image
source method implemented in \textit{pyroomacoustics} \cite{scheibler2018pyroomacoustics}.
The additive noise consists of diffuse-like stationary babble noise generated using \cite{habets2008generating,adrian2017synthesis}.

In the output layer of the FC-DNNs we consider $C$=72 DoA classes, leading to an angular resolution of 5°. 
Since for the considered array geometry $\tau^{\delta}_{max}$=14, including a margin, the input layer size for FC$_{full}$, FC$_{max}$, and FC$_{GA}$ is equal to 280, 10 and 20, respectively.
Training as well as evaluation are performed on non-overlapping Hann windowed 32\,ms signal frames at a  sampling rate of $f_{s}$=8\,kHz, yielding a frame size of 256\,samples.

\vspace*{-2mm}
\subsection{Training}
\label{sec:training}

\begin{figure}[t]
    \centering
    \includegraphics[width=0.7\linewidth]{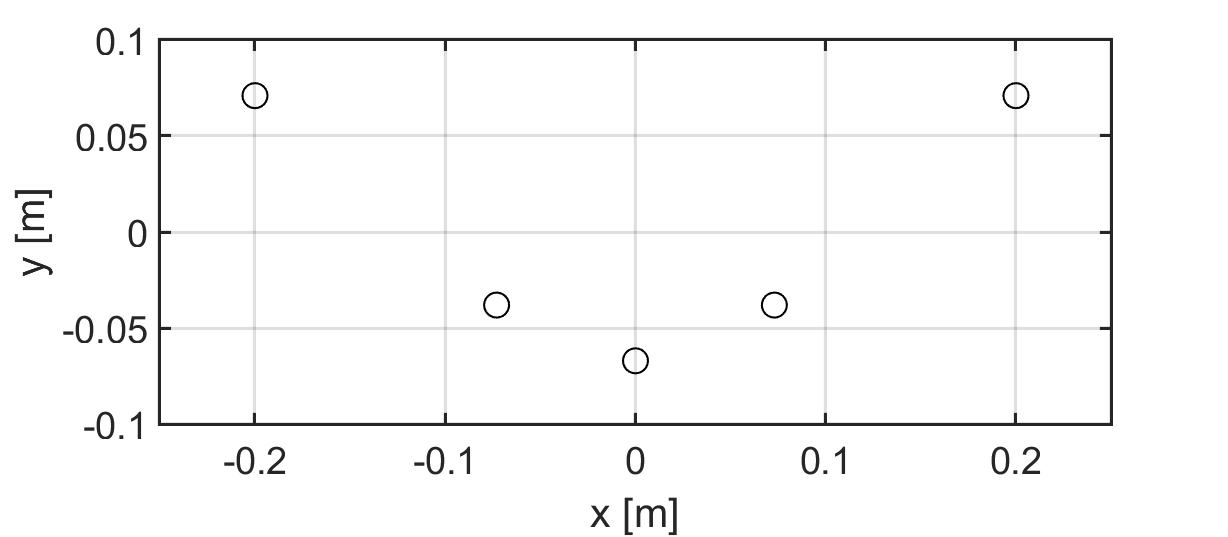}
    \vspace*{-4mm}
    \caption{2-dimensional arc-shaped microphone array geometry}
    \label{fig:array_shape}
    \vspace{-5mm}
\end{figure}

Every training sample consists of a single frame containing a single acoustic source as well as noise. We introduce variability in the acoustic parameters defining the training data aiming to achieve a robust algorithm that generalizes well to unmatched acoustic conditions. To this end, for every training sample a rectangular room with different dimensions and acoustic properties and an acoustic source with different directions and distances according to Table\,\ref{tab:Variations} is considered. 
FC$_{full}$, FC$_{max}$ and the CNN are trained using the microphone array geometry in Fig.\,\ref{fig:array_shape}.
The proposed geometry-aware FC$_{GA}$ is trained using a 2-dimensional array of randomly positioned microphones with a width and depth of 0.4\,m, whose coordinates are perfectly known, where a completely different array is considered for every training sample.
We employ mini batches of 32 samples, the Adam optimizer with a learning rate of $10^{-4}$, and a cross entropy loss function. Training is concluded when no reduction of the validation loss can be observed for 10 epochs.

\vspace*{-1mm}
\subsection{Performance Metrics}
\label{sec:performance_metrics}

Two performance metrics are used to quantify the performance of the considered algorithms: mean absolute error (\textit{MAE}) and \textit{Accuracy}. In order to account for circular wrapping, first the absolute angular error, $\delta_{n}$, is calculated for every estimate $n$ as
\vspace*{-1mm}
\begin{equation}
    \delta_{n} = \Big|\ \arg\left\{ e^{i2\pi\ \cdot\ (\tilde{\theta}_{n}-\theta_{n})\ /\ 360^{\circ}}\right\}\ \Big|\ \cdot\ \frac{360^{\circ}}{2\pi},
    \label{eq:Angular_Error}
\end{equation}
where $\tilde{\theta}_{n}$ is the estimated global DoA defined in Section \ref{sec:dnn_architecture} and $\theta_{n}$ is the ground truth DoA. Based on $\delta_{n}$ both performance metrics are then calculated as
\vspace*{-3mm}
\begin{eqnarray}
    \textit{MAE}\,\text{[°]} &=& \frac{1}{N}\sum\limits_{n=1}^{N}\delta_{n},\\
    \textit{Accuracy}\,\text{[\%]} &=& \frac{1}{N}\sum\limits_{n=1}^{N}\Theta\left(\varepsilon-\delta_{n}\right)\times100,
    \vspace*{-6mm}
\end{eqnarray}
where $N$ is the overall number of trials, $\Theta$ represents the Heaviside step function and $\varepsilon$ denotes the margin of tolerance -- here we use a margin of one DoA class, i.e. 5°.

\vspace*{-1mm}
\subsection{Evaluation Results}
\label{sec:evaluation}


This section presents the results of two experiments employing signals of length 5\,s with moderate reverberation and low noise.\vspace*{-1mm}\\
\\
\textit{First experiment: Coordinates deviating from trained geometry}\\
\vspace*{-1mm}
\\
The first experiment investigates the sensitivity of the baseline supervised learning-based algorithms (CNN, FC$_{full}$, FC$_{max}$) to coordinates deviating from the trained array geometry compared to the sensitivity of the proposed geometry-aware algorithm ($FC_{GA}$) and the performance of the baseline model-based algorithms.
In this experiment, random rooms and source positions are simulated according to Table\,\ref{tab:Variations}, but with a fixed reverberation time of $T_{60}$=0.5\,s and SNR=20\,dB.
Starting from the microphone array in Fig.\,\ref{fig:array_shape}, we consider microphone coordinate deviations in all directions. For every deviation in steps of $10^{-2}$\,m, all microphone coordinates deviate by the same amount, but in separate, random directions. Each deviation step is simulated $10^{4}$ times.

\begin{figure}[t]
\vspace*{-3mm}
    \centering
    \includegraphics[width=\linewidth]{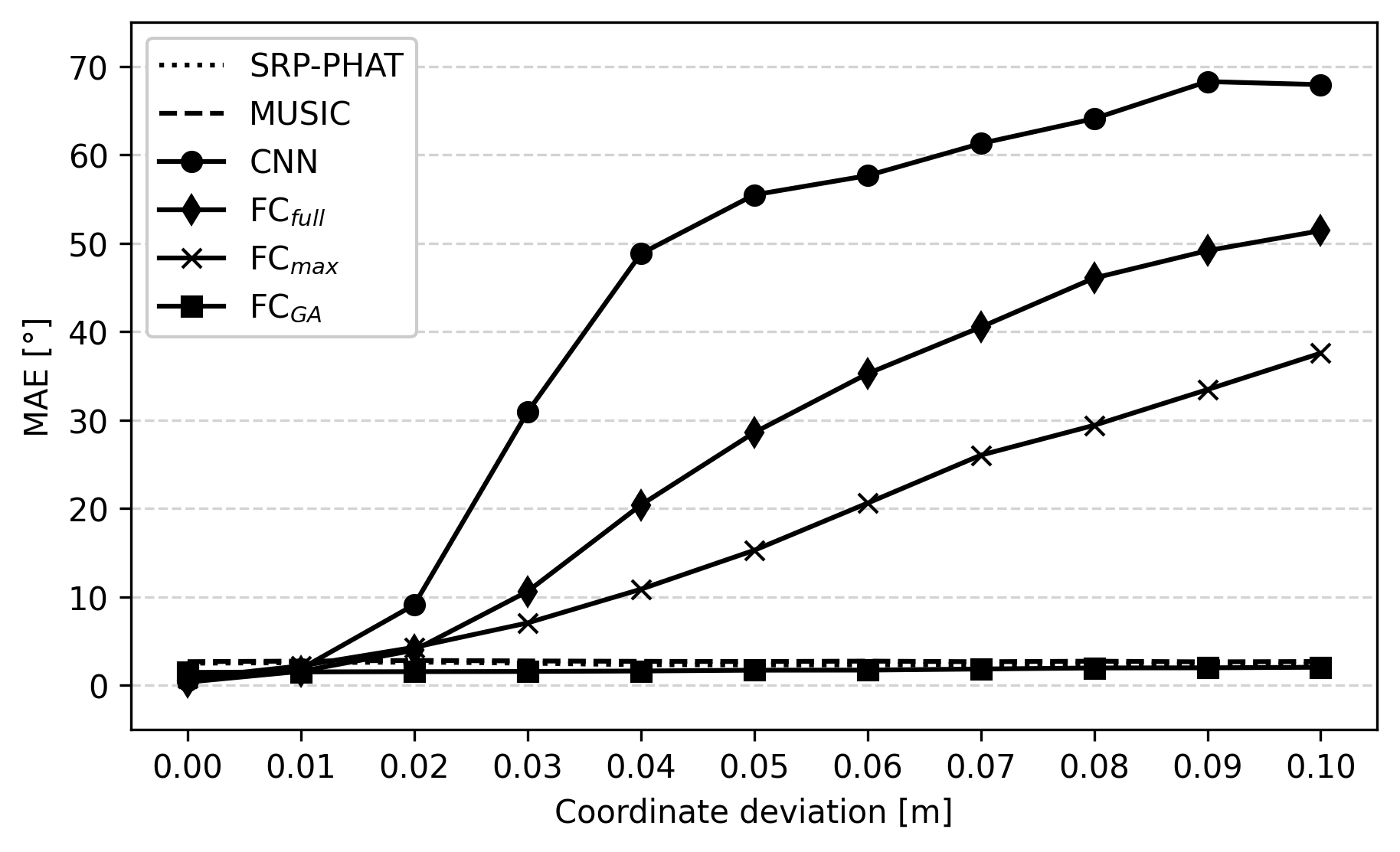}
    \vspace*{-8mm}
    \caption{MAE averaged over all simulations of model-based (SRP-PHAT, MUSIC) and supervised learning-based (CNN, FC$_{full}$, FC$_{max}$, FC$_{GA}$) algorithms with respect to deviating array coordinates ($T_{60}$=0.5\,s, SNR=20\,dB)}
    \vspace*{-1mm}
    \label{fig:Algos_coordinates_change_mae}
    \vspace*{-2mm}
\end{figure}

\begin{figure}[t]
    \centering
    \includegraphics[width=\linewidth]{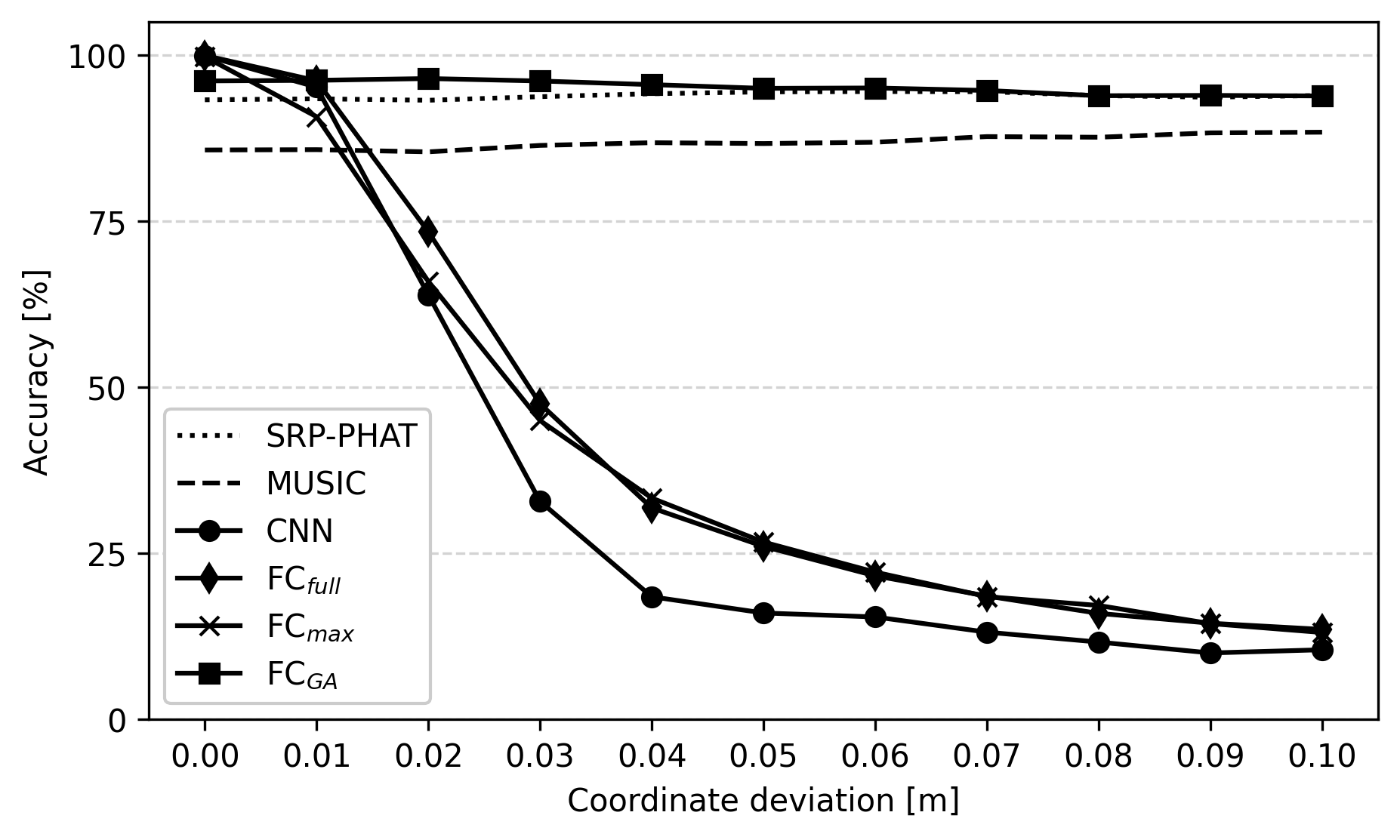}
    \vspace*{-8mm}
    \caption{Accuracy of model-based (SRP-PHAT, MUSIC) and supervised learning-based (CNN, FC$_{full}$, FC$_{max}$, FC$_{GA}$) algorithms with respect to deviating array coordinates ($T_{60}$=0.5\,s, SNR=20\,dB)} 
    \vspace*{-1mm}
    \label{fig:Algos_coordinates_change_acc}
    \vspace*{-5mm}
\end{figure}

Figs.\,\ref{fig:Algos_coordinates_change_mae} and\,\ref{fig:Algos_coordinates_change_acc} illustrate the effect of microphone coordinate deviation on all considered DoA estimation algorithms in terms of MAE and Accuracy, respectively. 
First, it can be observed that the baseline supervised learning-based algorithms (CNN, FC$_{full}$, FC$_{max}$) outperform the baseline model-based algorithms (SRP-PHAT, MUSIC) when no coordinate deviations occur. Second, up to a deviation of about 0.01\,m, no considerable decrease in performance can be observed for the supervised learning-based baseline systems. Larger deviations, however, lead to a substantial decrease in performance. Third, it can be clearly observed that the proposed geometry-aware algorithm (FC$_{GA}$) is much more robust to coordinate deviations than the baseline supervised learning-based algorithms.\vspace*{-1mm}\\
\\
\textit{Second experiment: Fully randomized geometry}\\
\vspace*{-1mm}
\\
We conducted $10^{4}$ simulations, each using a different 2-dimensional array geometry with uniformly distributed coordinates, with a maximum width and depth of 0.4\,m. For all considered algorithms in this experiment, we assume that the microphone coordinates are perfectly known. All other acoustic parameters are the same as in the first experiment, again with $T_{60}$=0.5\,s and SNR=20\,dB.

Table\,\ref{tab:results_second} shows the results in terms of MAE and Accuracy averaged over all simulations. These results show that the proposed geometry-aware algorithm (FC$_{GA}$) not only generalizes to unseen array geometries, but also outperforms the model-based baseline algorithms (SRP-PHAT, MUSIC), both in terms of MAE and Accuracy.

\begin{table}[h!]
\vspace*{-3mm}
    \centering
    \begin{tabular}{|l | r| r|}
        \hline
        \textbf{Algorithm} & \textbf{MAE [°]} & \textbf{Accuracy [\%]}\\
        \hline
        SRP-PHAT & 2.44 & 93.5\\
        MUSIC & 2.69 & 86.0\\
        FC$_{GA}$ & \textbf{1.47} & \textbf{96.1}\\
        \hline
        \end{tabular}
        \vspace*{-2mm}
    \caption{\label{tab:results_second}MAE and accuracy of model-based algorithms (SRP-PHAT, MUSIC) and proposed geometry-aware (FC$_{GA}$) algorithm for the randomized geometry simulation with $T_{60}$=0.5\,s and SNR=20\,dB}
\end{table}

\vspace*{-9mm}
\section{Conclusions}
\label{sec:conclusions}


In this paper we presented a feasibility study on geometry-aware DNN-based DoA estimation. First, we demonstrated the sensitivity of common supervised learning-based techniques towards deviations from the array geometry employed during training. The proposed geometry-aware algorithm uses as a novel feature the locations of GCC-PHAT maxima alongside the microphone coordinates. Experimental results showed that the proposed geometry-aware algorithm outperformed the evaluated model-based algorithms for different array geometries. Further studies will investigate the generalization to 3-dimensional arrays, the performance in adverse acoustic conditions as well as the robustness to inaccuracies in the assumed microphone coordinates.



\bibliographystyle{IEEEbib}
\bibliography{strings,refs}

\begin{thebibliography}{10}

\bibitem{huang2000microphone}
Y.~Huang, J.~Benesty, and G.~W. Elko,
\newblock ``Microphone arrays for video camera steering,''
\newblock in {\em Acoustic Signal Processing for Telecommunication}, pp.
  239--259. Springer, 2000.

\bibitem{doclo2010acoustic}
S.~Doclo, S.~Gannot, M.~Moonen, and A.~Spriet,
\newblock ``Acoustic beamforming for hearing aid applications,''
\newblock in {\em Handbook on array processing and sensor networks}, vol.~9,
  pp. 269--302. Wiley Hoboken, NJ, USA, 2010.

\bibitem{huang2008time}
Y.~A. Huang, J.~Benesty, and J.~Chen,
\newblock ``Time delay estimation and source localization,''
\newblock in {\em Springer Handbook of Speech Processing}, pp. 1043--1063.
  Springer, 2008.

\bibitem{madhu2008acoustic}
N.~Madhu and R.~Martin,
\newblock ``Acoustic source localization with microphone arrays,''
\newblock in {\em Advances in Digital Speech Transmission}, pp. 135--170. Wiley
  UK, 2008.

\bibitem{knapp1976generalized}
C.~H. Knapp and G.~C. Carter,
\newblock ``The generalized correlation method for estimation of time delay,''
\newblock {\em IEEE Trans. on Acoustics, Speech, and Signal Processing}, vol.
  24, no. 4, pp. 320--327, 1976.

\bibitem{1457990}
G.~C. Carter,
\newblock ``Coherence and time delay estimation,''
\newblock {\em Proc. of the IEEE}, vol. 75, no. 2, pp. 236--255, 1987.

\bibitem{dibiase2000high}
J.~H. DiBiase,
\newblock {\em A high-accuracy, low-latency technique for talker localization
  in reverberant environments using microphone arrays},
\newblock PhD thesis, Brown University, 2000.

\bibitem{schmidt1986multiple}
R.~Schmidt,
\newblock ``Multiple emitter location and signal parameter estimation,''
\newblock {\em IEEE Trans. on Antennas and Propagation}, vol. 34, no. 3, pp.
  276--280, 1986.

\bibitem{grumiaux2022survey}
P.-A. Grumiaux, S.~Kiti{\'c}, L.~Girin, and A.~Gu{\'e}rin,
\newblock ``A survey of sound source localization with deep learning methods,''
\newblock {\em The Journal of the Acoustical Society of America}, vol. 152, no.
  1, pp. 107--151, 2022.

\bibitem{wang2018robust}
Z.~Q. Wang, X.~Zhang, and D.~Wang,
\newblock ``Robust speaker localization guided by deep learning-based
  time-frequency masking,''
\newblock {\em IEEE/ACM Trans. on Audio, Speech, and Language Processing}, vol.
  27, no. 1, pp. 178--188, 2018.

\bibitem{8651493}
S.~Chakrabarty and E.~A.~P. Habets,
\newblock ``Multi-speaker {DOA} estimation using deep convolutional networks
  trained with noise signals,''
\newblock {\em IEEE Journal of Selected Topics in Signal Processing}, vol. 13,
  no. 1, pp. 8--21, 2019.

\bibitem{zhang2019robust}
W.~Zhang, Y.~Zhou, and Y.~Qian,
\newblock ``Robust {DOA} estimation based on convolutional neural network and
  time-frequency masking,''
\newblock in {\em Proc. Interspeech}, Graz, AUT, 2019, pp. 2703--2707.

\bibitem{mack2022signal}
W.~Mack, J.~Wechsler, and E.~A.~P. Habets,
\newblock ``Signal-aware direction-of-arrival estimation using attention
  mechanisms,''
\newblock {\em Computer Speech \& Language}, vol. 75, 2022,
\newblock article 101363.

\bibitem{krause2021data}
D.~Krause, A.~Politis, and K.~Kowalczyk,
\newblock ``Data diversity for improving {DNN}-based localization of concurrent
  sound events,''
\newblock in {\em Proc. European Signal Processing Conference (EUSIPCO)},
  Dublin, IRL, 2021, pp. 236--240.

\bibitem{johansson2002speaker}
A.~Johansson, N.~Grbi{\'c}, and S.~Nordholm,
\newblock ``Speaker localisation using the far-field {SRP-PHAT} in conference
  telephony,''
\newblock in {\em Proc. IEEE International Symposium on Intelligent Signal
  Processing and Communication Systems (ISPACS)}, Penang, MYS, 2002.

\bibitem{zhang21e_interspeech}
S.~Zhang and X.~Li,
\newblock ``Microphone array generalization for multichannel narrowband deep
  speech enhancement,''
\newblock in {\em Proc. Interspeech}, Brno, CZE, 2021, pp. 666--670.

\bibitem{9746876}
T.~Yoshioka, X.~Wang, D.~Wang, M.~Tang, Z.~Zhu, Z.~Chen, and N.~Kanda,
\newblock ``Vararray: Array-geometry-agnostic continuous speech separation,''
\newblock in {\em Proc. IEEE International Conference on Acoustics, Speech and
  Signal Processing (ICASSP)}, Singapore, SGP, 2022, pp. 6027--6031.

\bibitem{smith1987parshl}
J.~O. Smith and X.~Serra,
\newblock ``{PARSHL}: An analysis/synthesis program for non-harmonic sounds
  based on a sinusoidal representation,''
\newblock in {\em Proc. International Computer Music Conference (ICMC)},
  Champaign/Urbana, IL, USA, 1987, International Computer Music Conference, pp.
  290--297.

\bibitem{zhang2008does}
C.~Zhang, D.~Flor{\^e}ncio, and Z.~Zhang,
\newblock ``Why does {PHAT} work well in low noise, reverberative
  environments?,''
\newblock in {\em Proc. IEEE International Conference on Acoustics, Speech and
  Signal Processing (ICASSP)}, Las Vegas, NV, USA, 2008, pp. 2565--2568.

\bibitem{scheibler2018pyroomacoustics}
R.~Scheibler, E.~Bezzam, and I.~Dokmani{\'c},
\newblock ``Pyroomacoustics: A python package for audio room simulation and
  array processing algorithms,''
\newblock in {\em Proc. IEEE International Conference on Acoustics, Speech and
  Signal Processing (ICASSP)}, Calgary, AB, CAN, 2018, pp. 351--355.

\bibitem{panayotov2015librispeech}
V.~Panayotov, G.~Chen, D.~Povey, and S.~Khudanpur,
\newblock ``Librispeech: an {ASR} corpus based on public domain audio books,''
\newblock in {\em Proc. IEEE International Conference on Acoustics, Speech and
  Signal Processing (ICASSP)}, Brisbane, QLD, 2015, pp. 5206--5210.

\bibitem{habets2008generating}
E.~A.~P. Habets, I.~Cohen, and S.~Gannot,
\newblock ``Generating nonstationary multisensor signals under a spatial
  coherence constraint,''
\newblock {\em The Journal of the Acoustical Society of America}, vol. 124, no.
  5, pp. 2911--2917, 2008.

\bibitem{adrian2017synthesis}
J.~A. Adrian, T.~Gerkmann, S.~van~de Par, and J.~Bitzer,
\newblock ``Synthesis of perceptually plausible multichannel noise signals
  controlled by real world statistical noise properties,''
\newblock {\em Journal of the Audio Engineering Society}, pp. 914--928, 2017.

\end{thebibliography}

\end{document}